\documentclass{article}    

\usepackage{amsmath}
\usepackage{amssymb}
\usepackage{epsfig}
\usepackage{subfigure}
\usepackage{graphicx}

\begin{document}                                                                                   
\title{Isolated large amplitude periodic motions of towed rigid wheels} 

\author{D. Tak\'acs$^{\dag}$, G. St\'ep\'an$^{\dag}$ and S. J. Hogan$^{\ddag}$}

\maketitle
\begin{center}
\noindent{\it $^{\dag}$Department of Applied Mechanics, Budapest University of Technology and Economics}
\end{center}

\maketitle
\begin{center}
\noindent{\it $^{\ddag}$Department of Engineering Mathematics, University of Bristol}
\end{center}

\date{}
\maketitle

\begin{abstract}
This study investigates a low degree-of-freedom (DoF) mechanical model of shimmying wheels.
The model is studied using bifurcation theory and numerical continuation.
Self-excited vibrations, that is, stable and unstable periodic motions of the wheel, are detected with the help
of Hopf bifurcation calculations. These oscillations are then followed over a large parameter range for different
damping values 
by means of the software package AUTO97. For certain parameter regions, the branches representing large
amplitude stable
and unstable periodic motions become isolated following an isola birth. These regions are extremely
dangerous from an engineering view-point if they are not identified and avoided at the design stage.
\end{abstract}

\section{Introduction}  
Shimmy is a common name for the lateral vibration of a towed wheel. This has been a well-known phenomenon 
in vehicle systems dynamics for several decades: the name shimmy comes from a dance that was popular in the 
1930s.
One of the early scientific studies of shimmy dates back to this time \cite{Schlippe}.
In many cases, the appearance of shimmy is a serious problem, for example, in case of nose gears of airplanes 
or front wheels of motorcycles.
There are many mechanical models (see, for example, \cite{Pacejka1,Sharp,Plaut,OReilly,Troger,Leine})
that describe the shimmy of rolling wheels.
The two most important considerations of any model involve whether the wheel is rigid or elastic
and whether the suspension system is rigid or elastic. The simplest combination of a rigid wheel
and a rigid suspension system does not, by definition, support lateral vibrations.
Lateral vibration can occur in models with an elastic wheel and a rigid suspension (see \cite{Stepan})
and in models where both
the wheel and the suspension system are elastic.
But in this paper we wish to make analytic progress in order to obtain a clear understanding of the dynamics
and so we consider a low degree of freedom (DoF) mechanical model of a rigid wheel, which has a viscously damped 
elastic suspension. This model was studied without damping in \cite{Stepan1},
where subcritical Hopf bifurcations and chaotic and transient chaotic oscillations were found.

This paper is structured as follows. First, in Section 2, the mechanical model is introduced,
together with the equations of motion. In Sections 3 and 4, 
the Hopf bifurcation calculation is presented in the
presence of viscous damping at the suspension. The effect of damping on the stability of
stationary rolling is analysed. 
In Section 5, the periodic solutions of the system are followed using AUTO97 \cite{AUTO}, the 
stability
charts and the bifurcation diagrams are plotted and also compared to available analytical results.

The bifurcation
diagrams show an isola birth where isolated large amplitude stable and unstable periodic motions coexist with
the stable stationary rolling solution. These motions
are difficult to detect either by numerical simulation or by conventional stability and bifurcation
analysis. The presence of unstable periodic motions indicates a dangerous system configuration.

\section{Mechanical model}

\begin{figure}[H]
\centerline{\includegraphics[width=25pc]{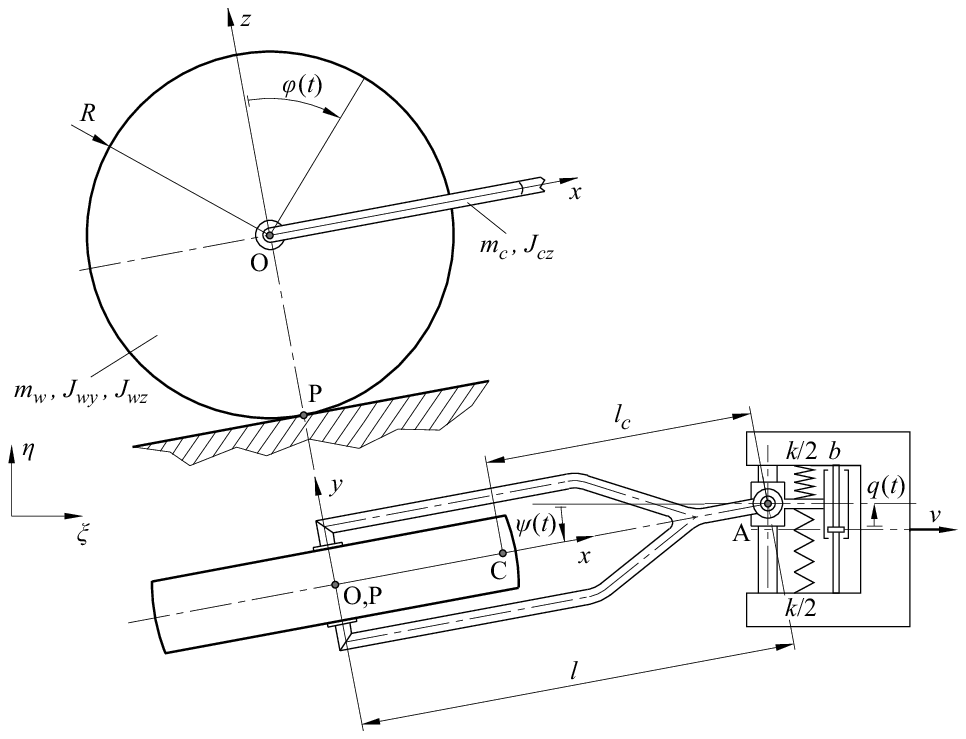}}
\caption{Model of rigid tyre.}\label{fi:model}
\end{figure}

The mechanical model under consideration is shown in Figure \ref{fi:model}. The plane of the rigid wheel is always vertical
to the ground, and they have a single contact point at P. The radius of the wheel is $R$, the mass of the wheel
is $m_{w}$ and its mass moment of inertia with respect to the $z$ axis is $J_{wz}$. The mass moment of inertia
with respect to the $y$ axis of its rotation is $J_{wy}$, where subscripts $w$ refers to wheel. 
The caster length is $l$, the distance of the center of gravity C of 
the caster and the king pin at A is $l_{c}$. The mass of the caster is $m_{c}$ and the mass moment of inertia with
respect to the $z$ axis at C is $J_{cz}$, with subscripts $c$ referring to caster. The system is towed in the
horizontal plane with constant velocity $v$. The king pin is supported by lateral springs of overall stiffness $k$
and viscous damping coefficient $b$.

Without rolling contraints, the system has 3 degrees of freedom, so one can choose 
the caster angle $\psi$, the king pin lateral position $q$, and the wheel rotation angle $\varphi$ as general coordinates.
The constraint of rolling (without sliding) means that the contact point P has zero velocity. This rolling
 condition leads to two scalar
kinematical constraint equations in the form of coupled first order nonlinear ordinary differential equations (ODEs) with respect
to the general coordinates.

The equations of motion of this rheonomic and non-holonomic system can be derived
with the help of the Routh-Voss equations or the Appell-Gibbs equations \cite{Gantmacher}. In the case of zero damping
($b=0$), they 
are given in \cite{Stepan2}. Here, we present the equations for non-zero viscous damping:

\begin{equation}
\begin{array}[l]{l@{\vspace{5pt}}}
\dot \psi = \vartheta \,,\\
\dot \vartheta  =  - \frac{{N\left( {\psi ,\vartheta ,q} \right)}}{{D\left( \psi  \right)}}\,, \\
\dot q = v\tan \psi  + \frac{l}{{\cos \psi }}\vartheta\,,\\
\dot \varphi  = \frac{{v + l\,\dot \psi \sin \psi }}{{R\cos \psi }}\,,
\end{array}
\end{equation}

\noindent where
\begin{equation}
\begin{array}{rcl}
 N\left( {\psi ,\vartheta ,q} \right) & = & \left( { - \left( {m_w l + m_c l_c } \right)v + \frac{{l{\kern 1pt} v}}{{R^2 }}J _{wy} \tan ^2 \psi  + \frac{{\left( {m_w  + m_c } \right){\kern 1pt} l{\kern 1pt} v}}{{\cos ^2 \psi }} + \frac{{b{\kern 1pt} l^2 }}{{\cos \psi }}} \right)\vartheta  \\ 
 & + & \left( {\left( {m_w  + m_c } \right)l^2  + \frac{{l^2 }}{{R^2 }}J _{wy} } \right)\frac{{\sin \psi }}{{\cos ^2 \psi }}\vartheta ^2  \\
 & + & k{\kern 1pt} l{\kern 1pt} q + b{\kern 1pt} l{\kern 1pt} {\kern 1pt} v\tan \psi  \\ 
 \end{array}
\end{equation}
and
\begin{equation}
\begin{array}{rcl}
D\left( \psi  \right) & = & \left( {m_c l_c \left( {l_c  - 2l} \right) - m_w l^2  + J _{wz}  + J _{cz}} \right)\cos \psi \\
& + & \left( { \frac{{\left( {m_w  + m_c } \right)l^2 }}{{\cos ^2 \psi }} + \frac{{l^2 }}{{R^2 }}J _{wy} \tan ^2 \psi } \right)\cos \psi 
\end{array}
\end{equation}

\noindent are odd and even functions respectively of the general coordinates. The first two equations in (1) are the
equations of angular momentum and the last two express the constraint of rolling (without sliding) of P.

Since the general coordinate $\varphi$ appears only in the fourth equation of motion, this coordinate is a so-called
cyclic one and the system can be described uniquely in the three dimensional phase space of the caster angle $\psi$,
caster angular velocity $\vartheta$ and the king pin lateral displacement $q$.

\section{Stability analysis}

The trivial solution of the system is stationary rolling along a straight line defined by the vector of the
towing velocity:
\\
\centerline{$\psi \equiv 0\,$, $\dot \psi \equiv 0\,$, $q \equiv 0$ and $\dot\varphi \equiv \frac{{v}}{{R}}\,$.}

When the towing speed is zero, the system forms a 1 DoF oscillator about the $z$ axis at P. The corresponding angular
natural frequency $\omega_n$ of the undamped linear system and the damping ratio $\zeta$ are given by

\begin{equation}
{\omega_n}  = \sqrt {\frac{{k\,l^2 }}{{J _{w2}  + J _c  + m_c (l - l_c )^2 }}}\,, \quad 
{\zeta} = \frac{1}{2}\frac{b}{k}{\omega_n}\,.
\end{equation}

\noindent Let us introduce the new dimensionless parameters:

\begin{equation}
\begin{array}{ccc}
L = \frac{l}{{l_c }}\,, & V = \frac{v}{{{\omega_n} \,l_c }}\,,
\end{array}
\end{equation}

\begin{equation}
\begin{array}{cc}
\kappa  = \frac{{m_c l_c (l - l_c )}}{{J _{w2}  + J _c  + m_c (l - l_c )^2 }}\,,
 & \chi  = \frac{{(m_c  + m_w )l^2  + J _{w1} {{l^2 } \mathord{\left/
 {\vphantom {{l^2 } {R^2 }}} \right.
 \kern-\nulldelimiterspace} {R^2 }}}}{{J _{w2}  + J _c  + m_c (l - l_c )^2 }}\,,
\end{array}
\end{equation}

\noindent where $L$ and $V$ are the dimensionless caster length and towing speed respectively, and $\kappa$ and $\chi$
are dimensionless mass moment of inertia parameters related to the caster and wheel geometry and inertia.

With these new parameters, the third order Taylor series expansion of the first three governing 
equations Eq. (1) about the trivial solution assumes the form:

\begin{equation}
\begin{array}{rcl}
\left[ {\begin{array}{*{20}c}
   {\dot \psi }  \\
   {\dot \vartheta }  \\
   {\dot q/l_c }  \\
\end{array}} \right] & = & \left[ {\begin{array}{*{20}c}
   0 & 1 & 0  \\
   { - 2{\zeta}V{\omega_n} ^2 /L} & { - {\omega_n} (2{\zeta} + \kappa V)} & { - {\omega_n} ^2 /L}  \\
   {{\omega_n} V} & L & 0  \\
\end{array}} \right]\left[ {\begin{array}{*{20}c}
   \psi   \\
   \vartheta   \\
   {q/l_c }  \\
\end{array}} \right]  \\
& + & \left[ {\begin{array}{*{20}c}
   0  \\
   { - {\omega_n} \left( {(2{\zeta} + \kappa V)(1 - \chi ) + V\left( {\frac{\chi }{L} - \frac{\kappa }{2}} \right)} \right)\psi ^2 \vartheta}   \\
   {\frac{{{\omega_n} V}}{3}\psi ^3  + \frac{L}{2}\psi ^2 \vartheta }  \\
\end{array}} \right] \\
& + & \left[
{\begin{array}{*{20}c}
{0} \\{- \chi \,\psi \,\vartheta ^2  - \frac{{{\omega_n} ^2 }}{L}\left( {\frac{1}{2} - \chi } \right)\psi ^2 \frac{q}{{l_c }} - \frac{{2{\zeta}V{\omega_n} ^2 }}{L}\left( {\frac{5}{6} - \chi } \right)\psi ^3 } \\
{0} \\
\end{array}} \right] \,.
\end{array}
\end{equation}

\noindent The characteristic equation is obtained from the linear coefficient matrix of Eq. (7):

\begin{equation}
  {\lambda}^3+\left(2\,{\zeta} + \kappa\,V\right)\,{\omega_n}\,{\lambda }^2+
\left(1 +\frac{2\,{\zeta}\,V}{L}\right)\,{{\omega_n} }^2\,\lambda+
\frac{V}{L}\,{{\omega_n}}^3=0 \,.
\end{equation}

\begin{figure}[H]
\centerline{\includegraphics[width=28pc]{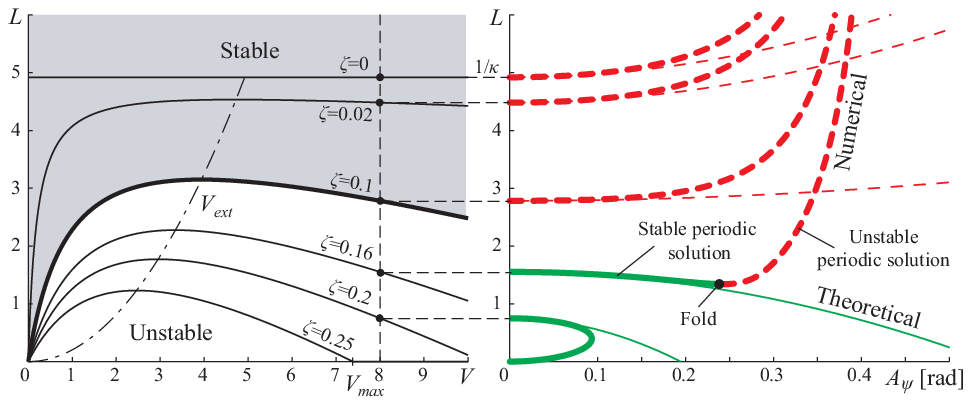}}
\centerline{\begin{tabular}{c@{\hspace{3pc}}c}
a.~~ The linear stability chart & b.~~ Bifurcation branches
\end{tabular}}
\caption{The stability chart and the bifurcation branches of the dimensionless system}\label{fi:dimlinstab}
\end{figure}

According to the Routh-Hurwitz criterion, the stability of the stationary rolling is equivalent to:

\begin{equation}
L \ge\,L_{cr}(V)  = \frac{{V(1 - 4{\zeta}^2  - 2{\zeta}\kappa V)}}{{2{\zeta} + \kappa V}} \,,
\end{equation}

\noindent considering positive parameter values only.
At fixed damping ratios $\zeta$,
the stability boundary curves are characterized by

\begin{equation}
V_{\rm{ext}}=\frac{1-2\zeta}{\kappa}\,, \quad V_{\rm{max}}=\frac{1-4\zeta^2}{2 \zeta \kappa}\,,
\end{equation}

\noindent where ${{\rm{d}}L_{cr}(V)}/{{\rm{d}}V}=0$ at $V=V_{\rm{ext}}$ and $L_{cr}(V)=0$ at $V=V_{\rm{max}}$.
Stationary rolling is always stable for $V>V_{\rm{max}}$ or $L>L_{\rm{cr}}(V_{\rm{ext}})$.
Figure \ref{fi:dimlinstab}.a shows the corresponding stability chart
in the plane of the dimensionless towing speed $V$ and caster length $L$ for $\kappa=0.203$ and $\chi=5.67$.
The stability region is shaded for $\zeta=0.1$. The dot-dash line is the locus of $V_{\rm{ext}}$ as a funcion of $\zeta$.
Our parameters come from a realistic towed wheel of a shopping trolley. The parameters of the wheel
are $m_w  = 0.3519\,\mathrm{[kg]}$, $J_{wy}=4.63 \cdot 10^{-5}\,\mathrm{[kgm^2]}$,
$J_{wz}=2.38 \cdot 10^{-5}\,\mathrm{[kgm^2]}$ and $R=0.04\,\mathrm{[m]}$. The data of the caster are
$m_c  = 0.0668\,\mathrm{[kg]}$, $l_c  = 0.012\,\mathrm{[m]}$ and $J_{cz}=3.48 \cdot 10^{-6}\,\mathrm{[kgm^2]}$.
We shall discuss Figure \ref{fi:dimlinstab}.b in Section 4.

\begin{figure}[h]
\centerline{\includegraphics[width=18pc]{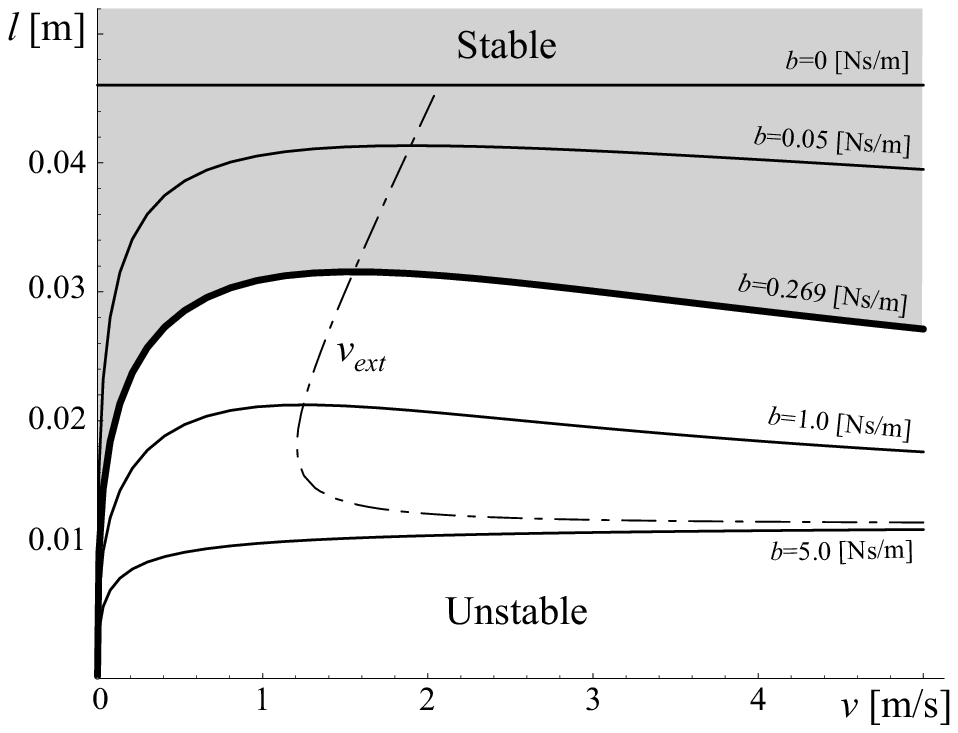}}
\caption{Stability boundaries of the linear system in dimensional form using
the shopping trolley parameters.}\label{fi:linstab}
\end{figure}

The stability chart in dimensional terms is of course of great practical use to engineers. However the natural
choice of scaling in Eq. (5) makes it difficult to immediately deduce its form from Figure \ref{fi:dimlinstab}.a.
Hence we redraw this Figure in dimensional terms in Figure \ref{fi:linstab}, using the shopping trolley parameters.

\section{Hopf bifurcation}

Despite giving a more useful stability chart, the dimensional form is less amenable to analysis. 
In dimensionless form the stability calculations are simpler and clearer.

The eigenvalues of the linear coefficient matrix of the dimensionless system can be determined on the stability boundary
when $L=L_{cr}$ as given in Eq. (9).
There are two complex conjugate eigenvalues with zero real part and one negative real eigenvalue:

\begin{equation}
\begin{array}{cc}
\lambda _{1,2}  =  \pm i\omega\,,
 & \lambda _3  =  - {\omega_n} \,(2{\zeta} + \kappa V)\,,
\end{array}
\end{equation}
where
\begin{equation}
\omega  = \frac{{\omega_n} }{{\sqrt {1 - 4{\zeta}^2  - 2{\zeta}\kappa V} }}\,,
\end{equation}

\noindent hence, there is Hopf bifurcation on the stability boundary.
The eigenvectors $\left\{ {\bf{s}}_1, {\bf{s}}_2, {\bf{s}}_3 \right\}$ can also be determined:

\begin{equation}
\begin{array}{c@{\vspace{5pt}}}
{\bf{s}}_1  = {\bf{\bar s}}_2  = \left[ {\begin{array}{c@{\vspace{5pt}}}
   {{\omega ^2 (2{\zeta} + \kappa V)\left( {1 + \rm{i}\,\frac{\omega }{{\omega_n} }\,(2{\zeta} + \kappa V)} \right)}}  \\
   { - {\omega ^3 (2{\zeta} + \kappa V)\left( {\frac{\omega }{{\omega_n} }\,(2{\zeta} + \kappa V) - \rm{i}} \right)}}  \\
   {V({\omega_n} ^2  + \omega ^2 (2{\zeta} + \kappa V)^2 )}  \\
\end{array}} \right],  \\
{\bf{s}}_3  = \left[ {\begin{array}{c@{\vspace{7pt}}}
   { -1 }  \\
   {{\omega_n} (2{\zeta} + \kappa V)}  \\
   {2{\zeta}V}  \\
\end{array}} \right].
\end{array}
\end{equation}

\noindent The transformation matrix $\bf{T}$ can be constructed from the eigenvectors in the following way:

\begin{equation}
{\bf{T}} = \left[ {\begin{array}{*{20}c}
   {{\mathop{\rm Re}\nolimits} \,{\bf{s}}_1 } & {{\mathop{\rm Im}\nolimits} \,{\bf{s}}_1 } & {{\bf{s}}_3 }  \\
\end{array}} \right].
\end{equation}

Let us introduce new variables $\left[ x_1,\,x_2,\,x_3 \right]$ such that:

\begin{equation}
\left[ {\begin{array}{*{20}c}
   \psi   \\
   \vartheta   \\
   {q/l_c }  \\
\end{array}} \right] = {\bf{T}}\left[ {\begin{array}{*{20}c}
   {x_1 }  \\
   {x_2 }  \\
   {x_3 }  \\
\end{array}} \right].
\end{equation}

If we substitute this into Eq. (7) and we multiply the equation on the left with
the inverse of the transformation matrix, the Poincar\'e normal form
is calculated, which has the form

\begin{equation}
\begin{array}{rcl}
\left[ \begin{array}[c]{l@{\vspace{5pt}}}
 \dot x_1  \\ 
 \dot x_2  \\ 
 \dot x_3  \\ 
 \end{array} \right] & = & \left[ {\begin{array}{@{\hspace{20pt}}c@{\vspace{2pt}\hspace{40pt}}c@{\vspace{2pt}\hspace{10pt}}c@{\vspace{2pt}}}
   0 & \omega  & 0  \\
   { - \omega } & 0 & 0  \\
   0 & 0 & { - \left( {2\zeta  + \kappa V} \right)\omega }  \\
\end{array}} \right]\left[ \begin{array}{l@{\vspace{5pt}}}
 x_1  \\ 
 x_2  \\ 
 x_3  \\ 
 \end{array} \right]  \\
 & + & \left[ \begin{array}{c}
 \sum\nolimits_{\scriptstyle j + k = 3 \hfill \atop 
  \scriptstyle j,k > 0 \hfill} {a_{jk} x_1^j x_2^k  + ...}  \\ 
 \sum\nolimits_{\scriptstyle j + k = 3 \hfill \atop 
  \scriptstyle j,k > 0 \hfill} {b_{jk} x_1^j x_2^k  + ...}  \\ 
 ... \\ 
 \end{array} \right].
\end{array}
\end{equation}

Since the nonlinearities are symmetric (i.e. there are no second degree terms in the nonlinear part of the Poincar\'e 
normal form), the centre manifold is approximated by a second degree surface. Thus, the transformation of the nonlinear part
needs only the terms in $x_1$ and $x_2$. The terms in which $x_3$ appear can be neglected. The sense of the Hopf
bifurcation comes from the reduced form of Poincar\'e-Lyapunov parameter $\delta$ for the symmetric case (see \cite{Stepan3}):

\begin{equation}
 \delta  = \frac{1}{8}\left( 3a_{30}  + a_{12}  + b_{21}  + 3b_{03}  \right)\,, 
\end{equation}

\noindent so

\begin{equation}
\delta  =   \frac{{\omega _n (2\zeta  + \kappa V)^2 }}{{8\,V^2 }} \cdot \frac{{ \zeta  + \kappa V - (2\zeta  + \kappa V)\frac{{\omega ^2 }}{{\omega _n ^2 }}\left( {2 - \zeta {\kern 1pt} \kappa \,V + \chi \left( {\frac{{\omega ^2 }}{{\omega _n ^2 }} - 2} \right)} \right)}}{{\left( {1 + (2\zeta  + \kappa V)^2 \frac{{\omega _n ^2 }}{{\omega ^2 }}} \right)\left( {(2\zeta  + \kappa V)^2  + \frac{{\omega _n ^2 }}{{\omega ^2 }}} \right)}}\,.
\end{equation}

\begin{figure}[H]
\centerline{\includegraphics[width=20pc]{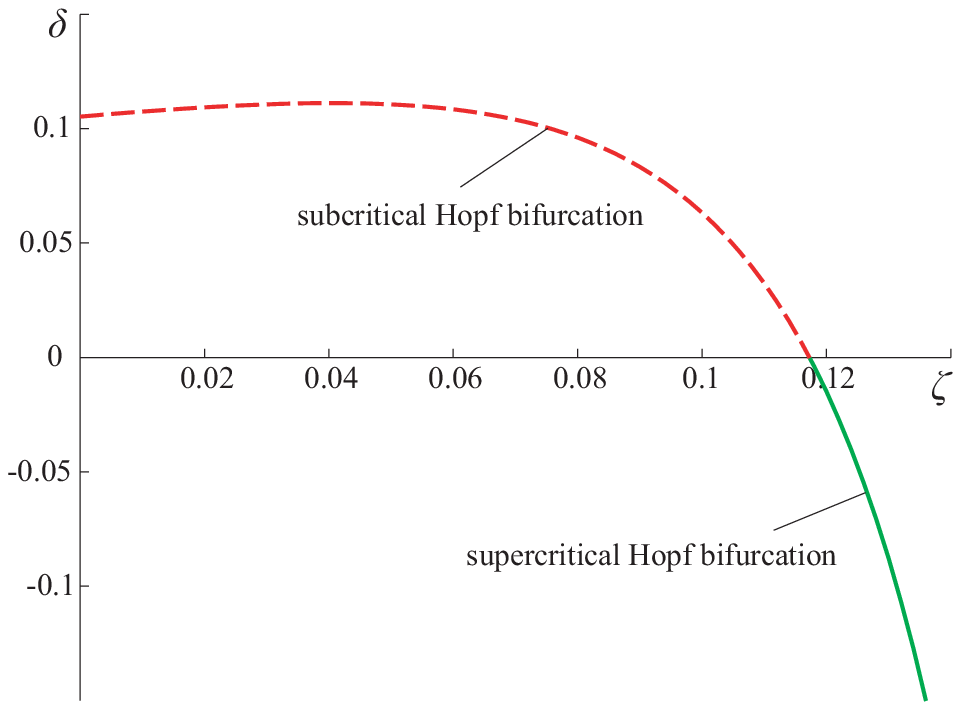}}
\caption{The $\delta$ parameter in Eq. (18) as a function of the damping $\zeta$.}\label{fi:delta}
\end{figure}

If $\delta$ is positive/negative then the periodic orbit of the Hopf bifurcation is subcritical/supercritical, namely
unstable/stable.
If we set $V=8$ and take $\kappa=0.203$ and $\chi=5.67$ as before, $\delta$ can be plotted as a function
 of the damping $\zeta$, see Figure \ref{fi:delta}. In our case the sign of $\delta$ changes at a critical 
value of the damping.

To compare our results with numerical continuation, the theoretical bifurcation branch has to be calculated. Accordingly
the real part of the implicit derivative of the characteristic equation with respect to the bifurcation parameter $L$ has
to be determined at the critical parameter value $L_{cr}$:

\begin{equation}
\left. {{\mathop{\rm Re}\nolimits} \frac{{{\rm{d}}\lambda }}{{{\rm{d}}L}}} \right|_{L = L_{cr} }  =  - \frac{{\omega _n (2\zeta  + \kappa V)^2 }}{{2V\left( {1 + (2\zeta  + \kappa V)^2 \frac{{\omega _n ^2 }}{{\omega ^2 }}} \right)}}\,.
\end{equation}

The amplitude of the periodic orbit is given by

\begin{equation}
r = \sqrt { - \frac{{{\mathop{\rm Re}\nolimits} \left. {\lambda '} \right|_{L = L_{cr} } }}{\delta }\left( {L - L_{cr} } \right)}\,,
\end{equation}

\noindent namely

\begin{equation}
r = \sqrt {\frac{{4V\left( {(2\zeta  + \kappa V)^2  + \frac{{\omega _n ^2 }}{{\omega ^2 }}} \right)}}{{\zeta  + \kappa V - (2\zeta  + \kappa V)\frac{{\omega ^2 }}{{\omega _n ^2 }}\left( {2 - \zeta {\kern 1pt} \kappa \,V + \chi \left( {\frac{{\omega ^2 }}{{\omega _n ^2 }} - 2} \right)} \right)}}\left( {L - L_{cr} } \right)}\,.
\end{equation}

\noindent With the help of the transformation matrix $\bf{T}$, the amplitude
of the vibration can be plotted with respect to the general coordinates $\psi$, $\vartheta$ and $q$. The theoretical branches
of the amplitude of $\psi$ ($\rm{A}_{\psi}$) are compared with the numerically continued results of the dimensionless system 
Eq. (7) in Figure \ref{fi:dimlinstab}.b. In this figure, as in the remainder of the paper, dashed lines mean
unstable branches and continuous lines mean stable solutions, and thick and thin lines show numerical and theoretical
results, respectively.
When the bifurcations are supercritical, the numerical continuation shows the existence of folds in certain parameter
regions, leading to large amplitude unstable periodic motions.
 
\begin{figure}[h]
\centerline{\includegraphics[width=20pc]{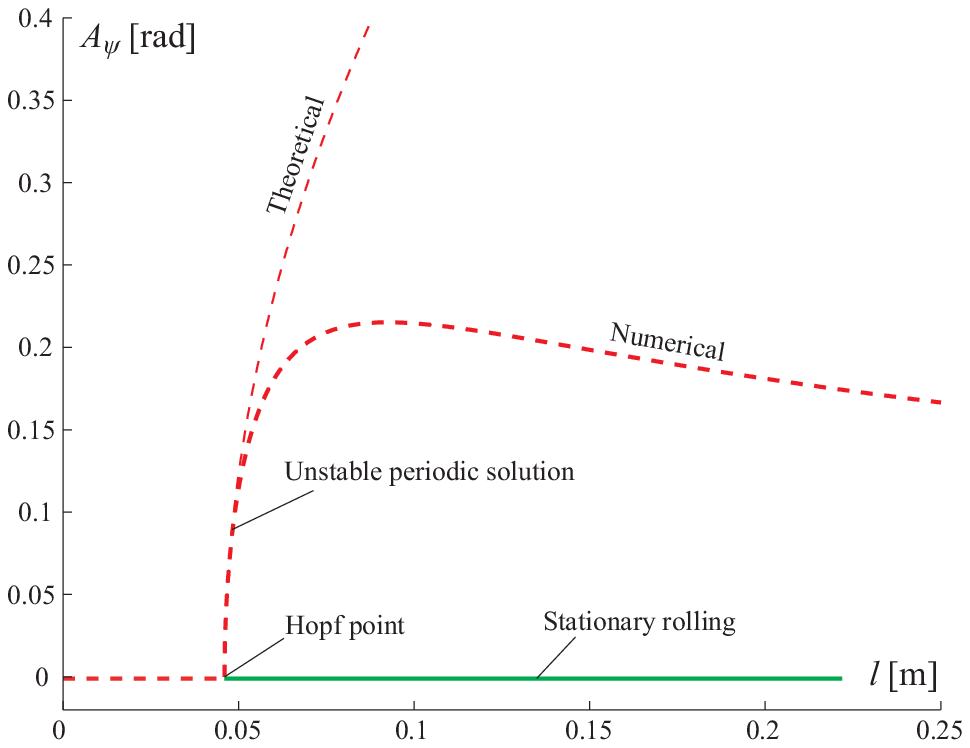}}
\caption{The theoretical and numerical bifurcation branches of the undamped system.}\label{fi:theo}
\end{figure}

As mentioned at the start of this section the analysis of the damped dimensional system is difficult. But in
case of zero damping the sense of the Hopf bifurcation, the Poincar\'e-Lyapunov parameter $\delta$, can be determined:

\begin{equation}
\delta  = \frac{{v\,\omega ^4 l}}{{8 {m_c (l-l_c) } \left( {v^2  + \omega ^2 l^2 } \right)^2 }}\left( {m_w l+ m_c l_c  + \frac{l}{{R^2 }}J _{wy} } \right)\quad  > 0 \,,
\end{equation}

\noindent which is always positive for all real parameter values, namely the Hopf bifurcation is always subcritical
(i.e. the periodic orbit is always unstable). The comparison of the theoretical and numerical results is shown in
Figure \ref{fi:theo}.

\section{Numerical Continuation}

\begin{figure}[h]
\centerline{\includegraphics[width=20pc]{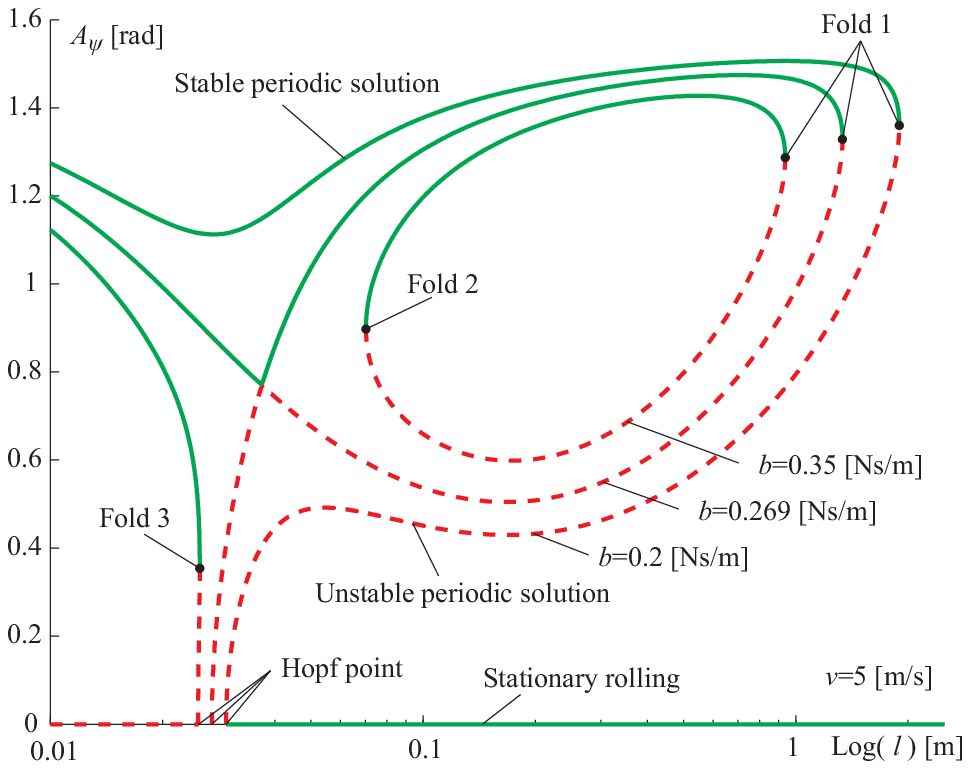}}
\caption{Bifurcation diagrams of the damped system with the isola birth.}\label{fi:isola}
\end{figure}

\begin{figure}[h]
\centerline{\includegraphics[width=20pc]{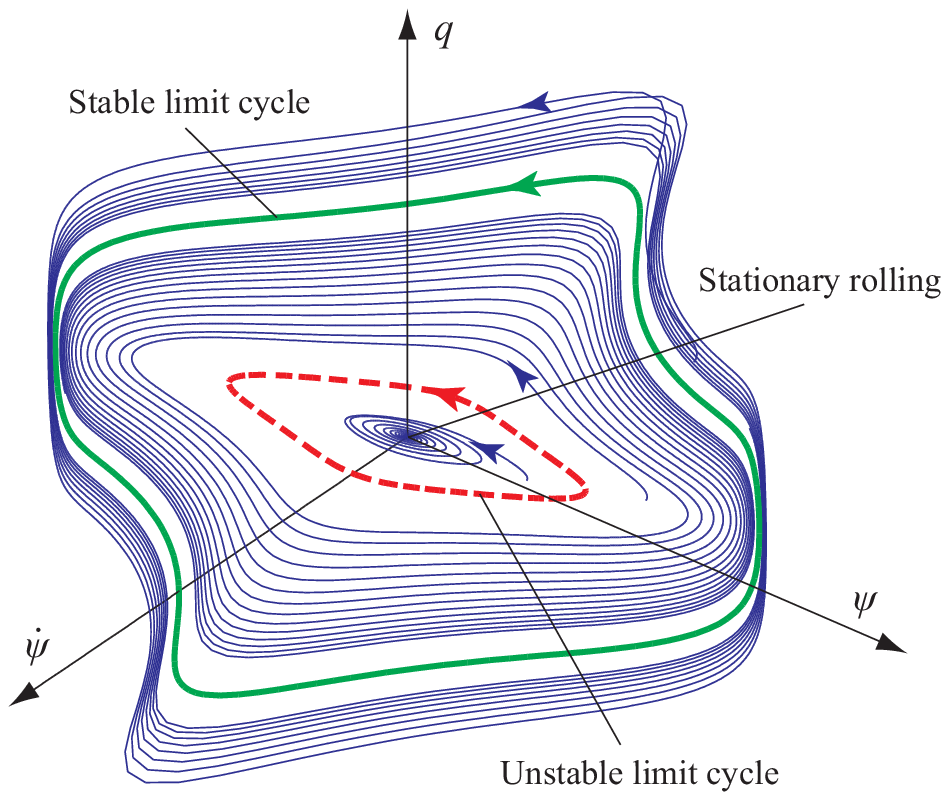}}
\caption{The trajectories in the phase space for $b=0.269\,\rm{[Ns/m]}$.}\label{fi:traj}
\end{figure}

Returning now to the damped case (with $b=0.2\,\rm{[Ns/m]}$), the branch of unstable periodic solutions was followed
using AUTO97 \cite{AUTO} from the Hopf point and a saddle-node bifurcation (fold) was detected,
see Figure \ref{fi:isola}. This means that
the damped system also has stable periodic solutions. So if the system is perturbed enough
then it will be attracted to these large amplitude stable periodic solutions.

\begin{figure}[h]
\centerline{\includegraphics[width=28pc]{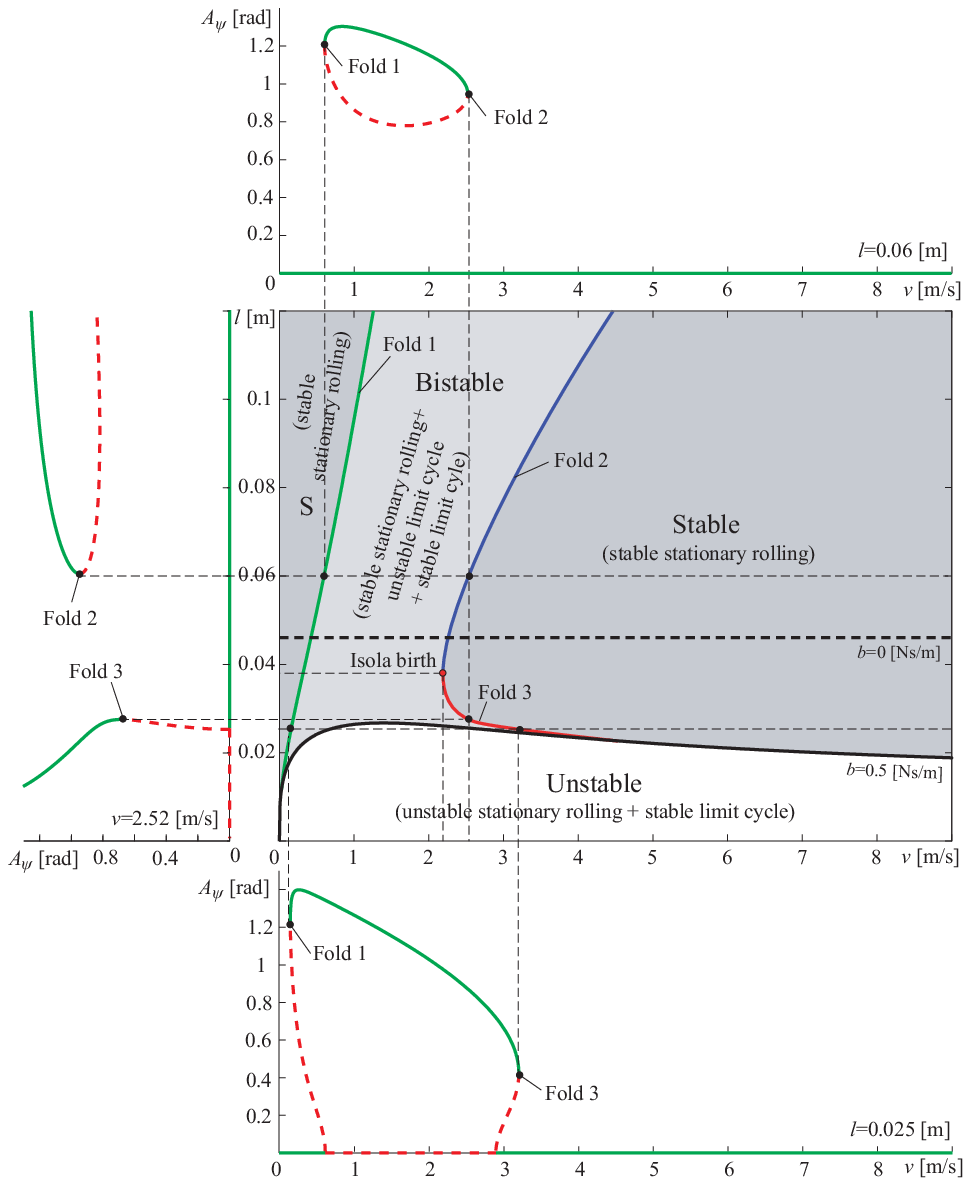}}
\caption{Stability chart of the nonlinear system.}\label{fi:stab}
\end{figure}

If the damping factor is increased, the peaks of the stable and unstable branches move closer together. At a
critical damping factor ($b=0.269\,\rm{[Ns/m]}$) the branches intersect each other and an isola is born,
see Figure \ref{fi:isola}.  Simulated motions are shown in phase space
in Figure \ref{fi:traj}. So a separated periodic
solution branch occurs in the bifurcation diagram for $b>0.269\,\rm{[Ns/m]}$. There are now three folds in the
 figure and, by changing the damping
factor $b$ or the towing velocity $v$, the location of these folds also changes.

In practice such large amplitude stable periodic motions might be expected to slide,
once a certain critical friction force is reached. Although we consider rolling, slipping is not part of our model
and this would have to be included if we wished to be certain that such motions could be observed in practice.

For fixed damping ratio a nonlinear stability chart can be plotted in the $(v,l)$ plane, where
the location of the folds can also be marked. With different projections, the nonlinear
behaviour of the system is represented in Figure \ref{fi:stab}.

\begin{figure}[h]
\centerline{\includegraphics[width=20pc]{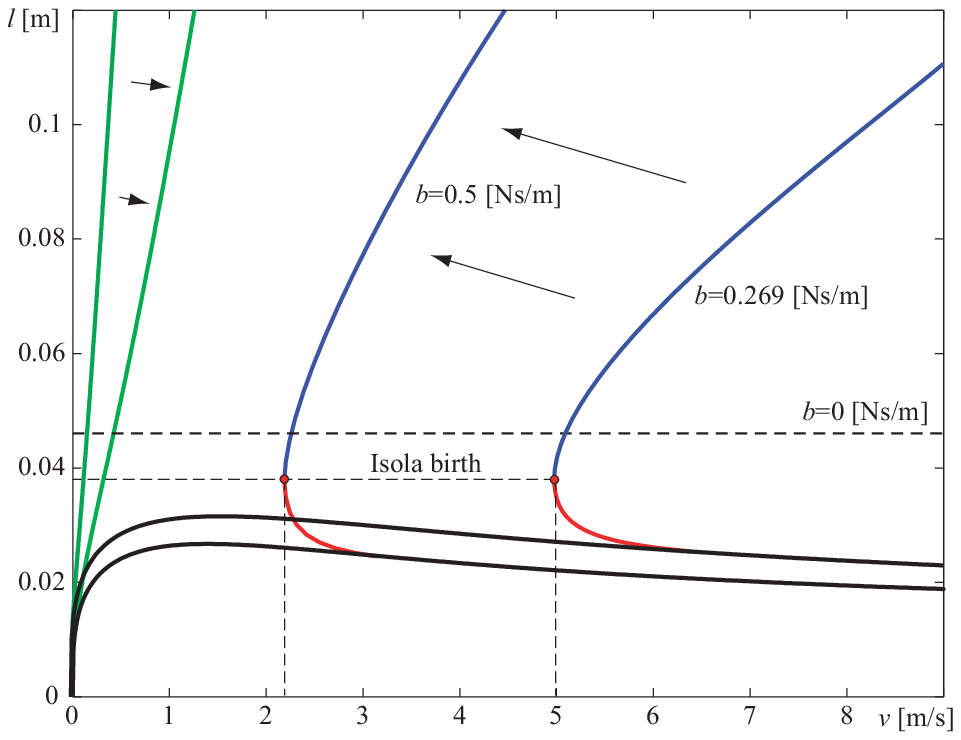}}
\caption{Stability boundaries at different damping factors.}\label{fi:stab12}
\end{figure}

The bistable area is of great importance. It has been found only with numerical continuation. That means that the
very dangerous large amplitude periodic motion can not be discovered analytically in this case. This bistable area
can be reduced and bounded for large enough damping ratio. This is shown in Figure \ref{fi:stab12}.

\section{Conclusions}

The importance in the investigation of the assumption of a rigid wheel model can be questioned, because a real system is more complex.
However the point of our study is to show that the well-known properties of shimmy, for example the existence of unstable
large amplitude periodic motions, are present even in this simple model.

In this paper the damped model was considered using analytical and numerical methods. It was shown that the 
subcritical Hopf bifurcation can change to supercritical if the damping ratio is increased. In the damped system 
a separated branch of periodic large amplitude solutions was detected using numerical continuation. For some parameter values,
the stationary rolling solution is stable for any value of the towing speed, but there is also a separated
periodic solution branch, a so-called isola of large amplitude, which is very dangerous.
This region can not be determined
by linear analysis. Therefore, without the nonlinear numerical investigation a system may be designed with parameter
values that allows for the presence of these dangerous solutions.

\subsection{Acknowledgements}
The authors greatly acknowledge the financial support provided by the Hungarian National Science 
Foundation under grant no.\ T043368.

\end{document}